\documentclass[%
superscriptaddress,
 amsmath,amssymb,
 aps,
prc,
]{revtex4-2}
\usepackage{graphicx}
\usepackage{dcolumn}
\usepackage{bm}
\usepackage{hyperref}
\usepackage[table]{xcolor}
\usepackage{array}
\hypersetup{breaklinks=true, colorlinks=true, citecolor=cyan, linkcolor=blue, urlcolor=blue,filecolor=blue}

\def\bea{\begin{eqnarray}}
\def\eea{\end{eqnarray}}

\begin{document}
\preprint{APS/PRC}

\title{Physical Implications of the Extrapolation and Statistical Bootstrap of Nucleon Structure Function Ratios $\frac{F_2^n}{F_2^p}$ for Mirror Nuclei $^3$He and $^3$H}

\author{Hannah~Valenty}%
\affiliation{Duquesne University, Pittsburgh, PA 15282, USA}%

\author{Jennifer~Rittenhouse~West}
\affiliation{Lawrence Berkeley National Laboratory, Berkeley, CA 94720, USA}

\author{Fatiha Benmokhtar}
\email{benmokhtarf@duq.edu}
\affiliation{Duquesne University, Pittsburgh, PA 15282, USA}%

\author{Douglas~W.~Higinbotham}
 \affiliation{Thomas Jefferson National Accelerator Facility, Newport News, VA 23606, USA}%

\author{Asia~Parker}
\affiliation{Duquesne University, Pittsburgh, PA 15282, USA}%

\author{Erin~Seroka}%
\affiliation{The George Washington University, Washington, DC 20052}%

\date{\today}

\begin{abstract}
A nuclear physics example of statistical bootstrap is used on the MARATHON nucleon structure function ratio data in the quark momentum fraction regions $x_B\rightarrow0$ and $x_B\rightarrow1$.  The extrapolated $F_2$ ratio as quark momentum fraction $x_B\rightarrow 1$ is $\frac{F_2^n}{F_2^p}\rightarrow 0.4 \pm 0.05$ and this value is compared to theoretical predictions.  The extrapolated ratio when $x_B\rightarrow 0$ favors the simple model of isospin symmetry with the complete dominance of seaquarks at low momentum fraction.  At high-$x_B$, the proton quark distribution function ratio $d/u$ is derived from the $F_2$ ratio  and found to be $d/u \rightarrow 1/6$.  Our extrapolated values for both the $\frac{F_2^n}{F_2^p}$ ratio and the $d/u$ parton distribution function ratio are within uncertainties of perturbative QCD values from quark counting, helicity conservation arguments and a Dyson-Schwinger Equation with contact interaction model.  In addition, it is possible to match the statistical bootstrap value to theoretical predictions by allowing two compatible models to act simultaneously in the nucleon wavefunction.  One such example is nucleon wavefunctions composed of a linear combination of a quark-diquark state and a 3-valence quark correlated state with coefficients that combine to give the extrapolated $F_2$ ratio at $x_B=1$.
\end{abstract}

\maketitle

\section{\label{sec:intro}Introduction}
The ratio of the deep inelastic structure functions $F_2^n / F_2^p$ provides fundamental information about the quark distributions of nucleons~\cite{Friedman:1991nq,Kendall:1991np}.
In particular, in the limit of the Bjorken scaling variable $x_B$~\cite{Bjorken:1968dy} going to unity, this ratio provides a powerful tool for discriminating between different partonic models.  Due to the lack of free neutron targets, it is very challenging to determine the ratio from scattering protons and neutrons; but using mirror nuclei,
${ }^3 \mathrm{H}$ and ${ }^3 \mathrm{He}$, and exploiting isospin symmetry allows for an accurate determination of this ratio.

The MARATHON experiment at Jefferson Lab has recently published their measured ratios of nucleon structure functions $F_2^n/F_2^P$ with deep inelastic scattering of electrons off of ${ }^3 \mathrm{H}$ and ${ }^3 \mathrm{He}$ nuclei \cite{JeffersonLabHallATritium:2021usd,Marathon:2021pro,mara1,Hague:2020dyx,mara3,mara4,mara5,mara6}.  Their results cover the quark momentum fraction range $0.19 < x_B < 0.83$ and significantly improve on previous measurements \cite{Breidenbach:1969kd,Bloom:1969kc,Friedman:1991nq}.  The high and low Bjorken-x regions were not covered and yet are necessary, especially in the $x_B\rightarrow1$ regime, in order to distinguish between theoretical predictions for quark and parton behavior at large momentum fraction.  The theory predictions differ in their interpretation of the components of the nucleon wavefunction and how they act in the high-$x_B$ regime.  One example is the scalar diquark model which predicts that at $x_B=1$ only the scalar $[ud]$ diquark plus valence quark component contributes, yielding a structure function ratio of $\frac{1}{4}$.  Another example is the $\rm SU(6)$ flavor model in which all vector diquarks (spin-1 and isospin-1 combinations) plus valence quark contribute to give a ratio of $\frac{2}{3}$.  Table \ref{tab:predict} presents an inexhaustive list of theoretical predictions, while theory reviews from 1996 \cite{Melnitchouk:1995fc} and 2010 \cite{Holt:2010vj} give comprehensive details and references.

Previous global analyses of parton distibution functions (PDFs) at next-to-next-to leading order (NNLO) by the CTEQ-TEA (Coordinated Theoretical/Experimental Project on QCD Phenomenology and Tests of the Standard Model) collaboration give a wide spread of $d/u$ parton distribution function (PDF) ratios \cite{Dulat:2015mca,Lai:1996mg}.  The extrapolation carried out in this work in the high-x region finds $d/u$ values differing from recent fits using the statistical Schlessinger point method, shown in Table \ref{tab:fits}.  Our value with uncertainties is within range of three of the theoretical models in Table \ref{tab:predict}.  However, it is possible that more than one theoretical model acts simultaneously in the high-$x_B$ region; for example, a quark-diquark configuration in the nucleon in linear combination with a 3-valence quark state can fit our extrapolated value.  We discuss this possibility in the next section.

We begin with a theory overview and analysis in the following section and then describe the extrapolation method in detail (Sec. \ref{sec:method}) before concluding.


\begin{table}
\begin{tabular}[c]{|c|c|c|}
\hline \textbf{Model} & \textbf{$F_2$ ratio} & \textbf{Ref.} \\
     & \textbf{prediction} &   \\
\hline Scalar diquark q[ud] & $1 / 4$ & \cite{Close:1973xw,Selem:2006nd} \\
\hline  Quark model/Isgur  & $1 / 4$ & \cite{Isgur:1998yb} \\
\hline  Holographic LFQCD  & $0.282$ & \cite{Liu:2019vsn,hlfqcd}\\
\hline  DSE with qq correlations  & $0.33$ & \cite{Bednar:2018htv}$^*$ \\
\hline  DSE-II contact interaction  & $0.41$ & \cite{Roberts:2013mja,Roberts:2011wy} \\
\hline pQCD helicity conservation & $3 / 7$ & \cite{Farrar:1975yb} \\
\hline Quark counting rules & 3/7 & \cite{Brodsky:1994kg} \\
\hline 3-quark correlations with CJ12MID & $\lesssim 0.51$ & \cite{Leon:2020bvt}$^*$ \\
\hline  DSE-I dressed quark masses  & $0.49$ & \cite{Roberts:2013mja} \\
\hline $\mathrm{SU}(6)$ Flavor & $2 / 3$ & \cite{Kuti:1971ph} \\ 
\hline 3-quark correlations with MMHT2014 & $\geq 0.68$ & \cite{Leon:2020bvt}$^*$ \\
\hline
\end{tabular}
\caption{$F_2^n / F_2^p$  Model Predictions as $x \rightarrow 1$.  An asterisk indicates that the reference gave a value for the valence quark ratio $d/u$, from which the $F_2$ is derived using the isospin symmetry assumptions $u^p = d^n =u$, $d^p=u^n=d$ and assuming the seaquark contribution is negligible.}
\label{tab:predict}
\end{table}

\begin{table}
\begin{tabular}[c]{|c|c|c|}
\hline \textbf{Method} & \textbf{$F_2$ ratio}  & \textbf{Reference} \\
\hline Schlessinger point method & $\approx 0.45$ & \cite{Cui:2021gzg}$^*$ \\
\hline CTEQ & unconstrained & \cite{Dulat:2015mca,Lai:1996mg} \\ 
\hline JAM & $\approx 0.4$ but $x\lesssim 0.8$ & \cite{Cocuzza:2021rfn} \\
\hline
\end{tabular}
\caption{$F_2^n / F_2^p$ fits as  $x \rightarrow 1$.  An asterisk indicates that the reference gave a value for the valence quark ratio $d/u$, from which the $F_2$ is derived using the isospin symmetry assumptions $u^p = d^n =u$, $d^p=u^n=d$ and negligible seaquark contributions.}
\label{tab:fits}
\end{table}

\vspace{-0.445cm}
\section{\label{sec:theory}Structure Function Theory}
We start with a review of relevant structure function definitions.  The internal structure of nucleons is encoded in the form of structure functions $F_2(x_B,Q^2)$ describing quark and antiquark behavior in terms of parton momentum distribution functions weighted by the square of their electric charge.

Structure functions are components of the deep-inelastic scattering differential cross-section for charged lepton scattering, {\it e.g.} ${\mathrm{eN} \rightarrow \mathrm{e} \mathrm{X}}$,
\begin{equation}
\frac{\mathrm{d}^{2} \sigma}{\mathrm{d} t \mathrm{~d} u}= \frac{4 \pi \alpha^{2}}{t^{2}} \frac{1}{2} \frac{1}{s^{2}(s+u)} \left[2 x F_{1}(s+u)^{2} - 2 u s F_{2}\right],
\end{equation}
given in terms of the Mandelstam variables for the 4-momenta of the incoming lepton $p_1$, target $p_2$, outgoing lepton $p_3$, and hadronized debris $p_4$,   $s=\left(p_{1}+p_{2}\right)^{2}$, 
$t=\left(p_{1}-p_{3}\right)^{2}$, 
$u=\left(p_{1}-p_{4}\right)^{2}$.  The nucleon structure function $F_2$ is defined (up to leading order in the strong force coupling $\alpha_S$) as
\begin{equation}
F_{2}(x_B)=2 x_B F_{1}(x_B)=\sum_{\mathrm{i}} Q_{\mathrm{i}}^{2} x (q_{i}(x_B)+ \bar{q}_{i}(x_B))
\end{equation}
where the sum is over quark flavors, $Q_{\mathrm{i}}^{2}$ is the charge on the $i$th flavor of quark, $q_{i}(x_B)$ are the quark momentum distribution functions, and Bjorken-x is the fraction of nucleon momentum carried by the struck quark.  Bjorken-x is given in terms of experimental variables
\begin{equation}
x_B=\frac{Q^{2}}{2 M_{\mathrm{\rm T}} \nu},
\end{equation}
where $\nu$ is the energy lost by the lepton, $E-E'$, $M_{\rm T}$ is typically the mass of the struck nucleon in the target, $Q^2$ is minus the square of the 4-momentum transfer (the virtual photon 4-momentum squared) $Q^{2} \equiv-q^{2}=2 E E^{\prime}(1-\cos \theta)$ where $\theta$ is the lepton scattering angle.  The subscript $B$ will be dropped for notational simplicity from here on out.

\begin{figure}[ht!]
    \centering
    \includegraphics[width =0.48\textwidth]{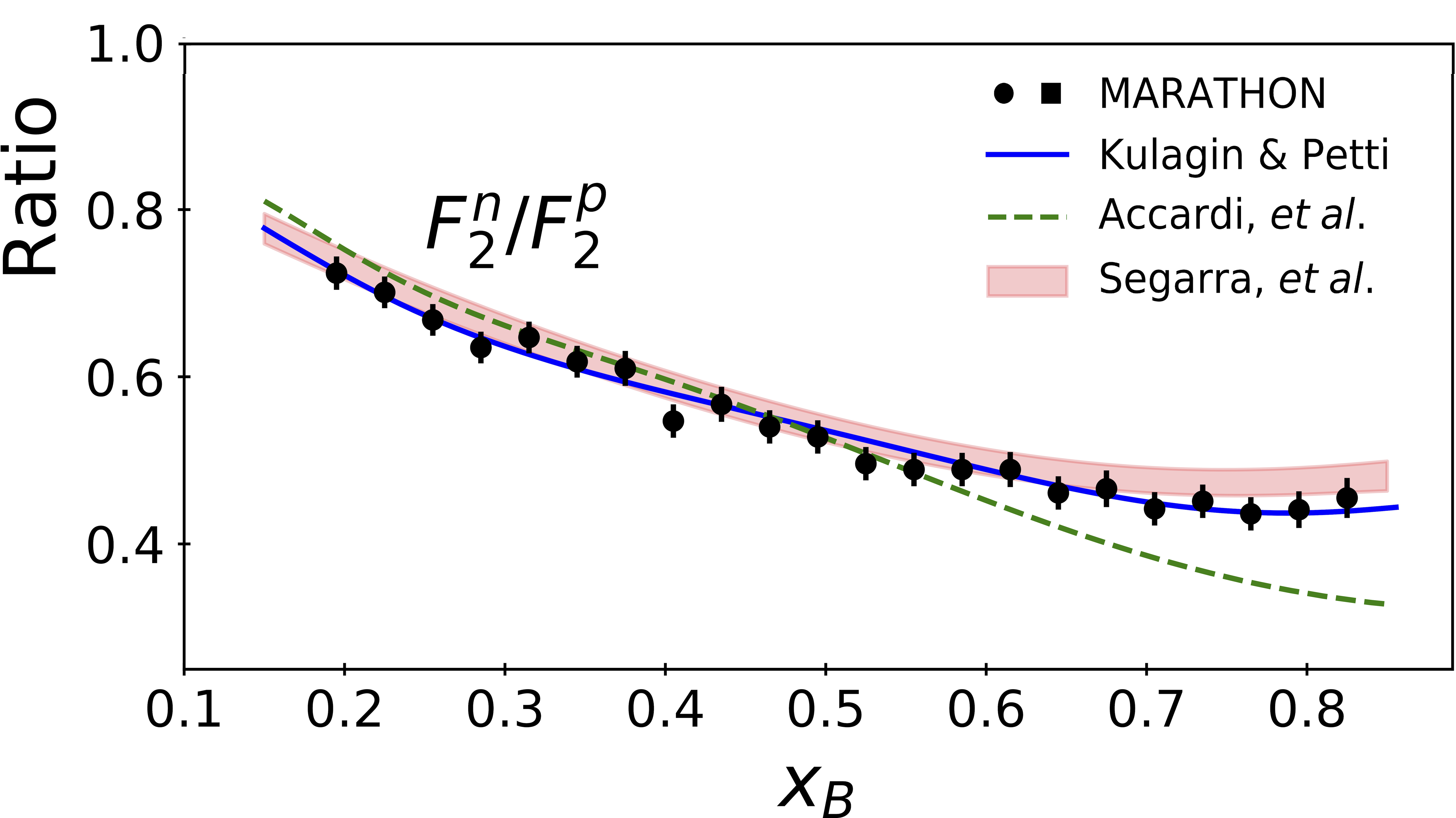}
    \caption[]{Deep inelastic scattering $F_2^n / F_2^p$ ratios vs. $x_B$ from MARATHON \cite{JeffersonLabHallATritium:2021usd} compared to theoretical predictions. Error bars include overall systematic uncertainties. All curves correspond to MARATHON kinematics. Plot credit: T.Kutz.}
    \label{marathon2}
\end{figure}

The ratio of neutron to proton structure functions $F_{2}^{n} / F_{2}^{p}$ as measured by the MARATHON experiment and in the regions extrapolated by the statistical bootstrap method on MARATHON data are shown in Fig.\ref{marathon2} and Fig.\ref{stat-bootfinal}, respectively.  Physical implications and simplifying assumptions of the low and high Bjorken-x behavior are discussed next.

\subsection{Isospin symmetry analysis of extrapolated behavior of $F_{2}^{n} / F_{2}^{p}$: $x\rightarrow0$ region}
\begin{figure}[ht!]
    \centering
    \includegraphics[width =0.4\textwidth]{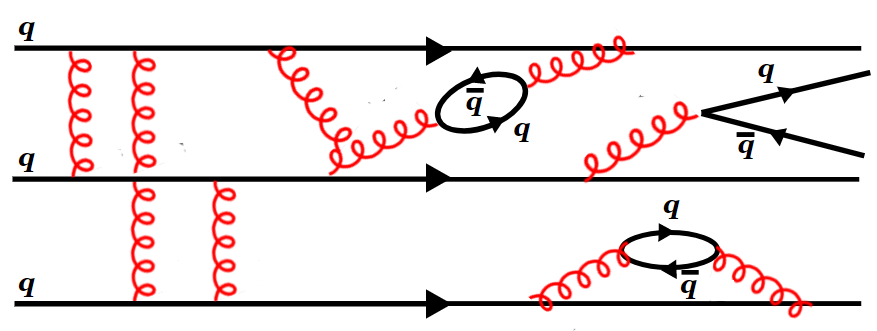}
    \caption[]{Example of low-x gluon splitting in a 5-quark Fock state of a nucleon. Adapted from  \cite{Close:1979bt}.}
    \label{fig:brem}
\end{figure}

In the low-x region, the process of gluon bremsstrahlung following by gluon splitting to quark-antiquark pairs dominates the nucleon structure function due to a $\frac{dk}{k}$ probability for gluon splitting,  Fig.\ref{fig:brem} \cite{Smallx:2002him}. The structure function in this region
\bea 
F_{2}(x)=\sum_{i} Q_{i}^{2} x (q_{i}(x) + \bar{q}_{i}(x))
\eea 
contains contributions from the light quark $q\bar{q}$ pairs, $u\bar{u}$, $d\bar{d}$, $s\bar{s}$.  Gluon splitting to the heavy mass quarks $c$, $b$, $t$ is neglected as these are highly virtual processes and contribute negligibly to the scattering amplitude.  

The structure functions are therefore written as 
\bea 
\begin{aligned}
\frac{1}{x} F_{2}^{p}(x)=&\frac{4}{9} u^{p}(x) + \frac{1}{9} d^{p}(x) +  \frac{4}{9}\left[u_s^{p}(x)+\bar{u}_s^{p}(x)\right] \\ &+\frac{1}{9}\left[d_s^{p}(x)+\bar{d}_s^{p}(x)\right]  
+\frac{1}{9}\left[s_s^{p}(x)+\bar{s}_s^{p}(x)\right] 
\end{aligned}
\eea 
and 
\begin{equation}
\begin{aligned}
\frac{1}{x} F_{2}^{n}(x)=&\frac{4}{9} u^{n}(x) + \frac{1}{9} d^{n}(x) + 
\frac{4}{9}\left[u_s^{n}(x)+\bar{u}_s^{n}(x)\right]\\
&+\frac{1}{9}\left[d_s^{n}(x)+\bar{d}_s^{n}(x)\right]+ \frac{1}{9}\left[s_s^{n}(x)+{\bar{s}_s}^{n}(x)\right]
\end{aligned}
\end{equation}
with subscript $s$ denoting sea quark distributions.

The following assumptions are now made for the quark momentum distribution functions.  Strong isospin symmetry assumes $u$ and $d$ form an $\rm SU(2)_{iso}$ doublet (as do $p$ and $n$) which implies the relations  \cite{Close:1979bt}
\begin{equation}
\begin{aligned}
&u^{p} = d^{n} ~\equiv u \\
&d^{p} = u^{n} ~\equiv d \\
&s^{p} = s^{n} ~\equiv s.
\end{aligned}
\label{eq:iso-assume}
\end{equation}
The rest of the analysis is therefore given in terms of the proton's parton distribution functions, the same approximations as used in MARATHON \cite{JeffersonLabHallATritium:2021usd}.

The structure functions can be rewritten as
\begin{equation}
\begin{aligned}
&\frac{1}{x} F_{2}^{p}=\frac{4}{9}u + \frac{1}{9}d + ~{\rm sea~terms} \\
&\frac{1}{x} F_{2}^{n}=\frac{4}{9} d + \frac{1}{9} u + ~{\rm sea~terms}.
\end{aligned}
\label{eq:structures}
\end{equation}

For simplicity, assume that the sea quark distributions are the same, as in \cite{Close:1979bt},
\begin{equation}
u_{\mathrm{s}}(x)=\bar{u}_{\mathrm{s}}(x)=d_{\mathrm{s}}=\bar{d}_{\mathrm{s}}=s_{\mathrm{s}}=\bar{s}_{\mathrm{s}} \equiv K.
\end{equation}
This assumption does not affect the results.

Then we have
\begin{equation}
\begin{aligned}
&\frac{1}{x} F_{2}^{n}=\frac{1}{9}\left(u+4 d\right)+\frac{12}{9} K \\
&\frac{1}{x} F_{2}^{p}=\frac{1}{9}\left(d+4 u\right)+\frac{12}{9} K.
\end{aligned}
\label{eq:f2-at-lowx}
\end{equation}
 When sea quark momentum distribution functions dominate the structure functions, $K\gg u, d$, we find 
\begin{equation}
\frac{F_{2}^{n}}{F_{2}^{p}}(x) \rightarrow 1~{\rm for}~x\rightarrow 0.
\end{equation}
This conclusion agrees with the extrapolated value shown in Fig.\ref{stat-bootfinal}.  Note that it does not depend on the assumption of equal sea quark distribution functions.  It requires only that the \textit{sum} of sea quark distribution functions to be the same for the proton and the neutron, which is essentially a sea quark version of the isospin symmetry assumptions of Eq. \ref{eq:iso-assume}.

\subsection{Isospin symmetry analysis of extrapolated behavior of $F_{2}^{n} / F_{2}^{p}$: $x\rightarrow 1$ region}
In the high-x region, for $x\rightarrow 1$, the extrapolated ratio $F_{2}^{n} / F_{2}^{p} \rightarrow 0.4 \pm 0.05$.  The sea quark distribution functions become negligible in this region of Bjorken-x and the structure functions are given by 

\begin{equation}
\begin{aligned}
&\frac{1}{x} F_{2}^{n}=\frac{1}{9}\left(u+4 d\right)\\
&\frac{1}{x} F_{2}^{p}=\frac{1}{9}\left(d+4 u\right)
\end{aligned}
\label{eq:f2-at-highx}
\end{equation}
where the symmetries of Eq.\ref{eq:iso-assume} are again assumed.

For this analysis, we set the ratio equal to the extrapolated value of $0.4$ and solve for the ratio of quark distribution functions which may then be compared to data.  The result is
\begin{equation}
\frac{F_{2}^{n}} {F_{2}^{p} } = \frac{\frac{1}{9}\left(u+4 d\right)}{\frac{1}{9}\left(d+4 u\right)}=0.4 \implies \frac{d}{u} = \frac{1}{6}
\end{equation}

This value is close to the lower limit of recent work \cite{Cui:2021gzg}, 
\begin{equation}
\frac{d}{u} \vert_{x \rightarrow 1}=0.230(57)
\end{equation}
and is due only to the isospin symmetry arguments of Eq.\ref{eq:iso-assume} and the vanishing of sea quark distributions at high-x.  Earlier theoretical work argued that a scalar $[ud]$ diquark forms within the nucleon in the $x_B\rightarrow 1$ limit as a way to lower the energy of the system \cite{Selem:2006nd} (Table \ref{tab:predict}).  In this version of the scalar diquark model, the highest probability state with a single quark containing all of the nucleon momentum must minimize the energy of the remaining constituents.  Therefore, as $x\rightarrow1$, the lowest energy proton state is that which has the $u$ quark carrying all the longitudinal momentum with the remaining valence quarks forming the spin-0 isospin-0 $[ud]$ diquark \cite{Jaffe:2004ph} with an estimated binding energy of nearly $150~\rm MeV$ \cite{West:2020tyo}.  Similarly for the neutron with momentum distribution functions swapped $u \leftrightarrow d$, which gives $d/u=0$ in the limit of $x\rightarrow 1$.  The structure function ratio in this case is given by 
\begin{equation}
\lim _{x \rightarrow 1} \frac{F_{2}^{n}(x)}{F_{2}^{p}(x)} =  0.25.
\end{equation}
Our analysis finds the limiting behavior 
\begin{equation}
    \lim_{x \rightarrow 1} \frac{F_{2}^{n}(x)}{F_{2}^{p}(x)} =  0.4 \pm 0.05
    \end{equation}
which disfavors this value.  We most closely match the perturbative QCD (pQCD) values using helicity conservation, hard gluon exchange \cite{Farrar:1975yb} and quark counting rules \cite{Brodsky:1994kg}.  The latter two models give the same ratio limit,
\begin{equation}
    \lim_{x \rightarrow 1} \frac{F_{2}^{n}(x)}{F_{2}^{p}(x)} =3/7 \approx 0.43,
    \end{equation}
and these are within the bootstrap uncertainties as shown by the blue band of Fig.\ref{stat-bootfinal}.  There is another physical implication, however, because while some models are incompatible with each other, it is possible that two compatible models could act over the same physical region to give the extrapolated value.  One such example is the scalar diquark model with $F_2^n/F_2^P\rightarrow 0.25$ \cite{Selem:2006nd} together with the 3-quark correlation plus MMHT2014NNLO with $F_2^n/F_2^P \geq 0.68$ (or CJ12MID with ratio $\lesssim 0.51$) from \cite{Leon:2020bvt}, with different weights given to scalar diquark vs. 3-quark correlation contributions in order to yield a final ratio of $0.4$.  The nuclear wave function for A=3 nuclei has been proposed as a linear combination of nucleons in quark-diquark and 3-valence quark configurations, with a range of predictions for short-range correlations \cite{West:2020tyo} which matches the data \cite{Li:2022fhh}.  We now move on to describing the extrapolation method.

\section{\label{sec:method}Statistical Bootstrap Method}
Bootstrapping is a technique that allows an estimation of the sampling distribution of almost any statistical distribution using the method of sampling data with replacement. It is any test or metric that falls under the broader class of re-sampling  methods. Bootstrapping assigns measures of accuracy such as variance, confidence interval  and prediction errors to sample estimates \cite{Efron:1993}.

In the case of the MARATHON experiment, a rational function $R_f$ of the form of
\begin{equation}
R_{f}=n_0 \frac{1+x n_1}{1+x m_1}
\end{equation}
was used to fit the 22 data points corresponding to $x_B$ values between  0.195 and 0.825. $n_0$, $n_1$ and $m_1$ are the three fit parameters of this function. The fit was then extrapolated to lower and higher limits of $x_B$ to cover the full range from 0 to 1. 

\begin{figure}
    \includegraphics[width =0.45\textwidth, ]{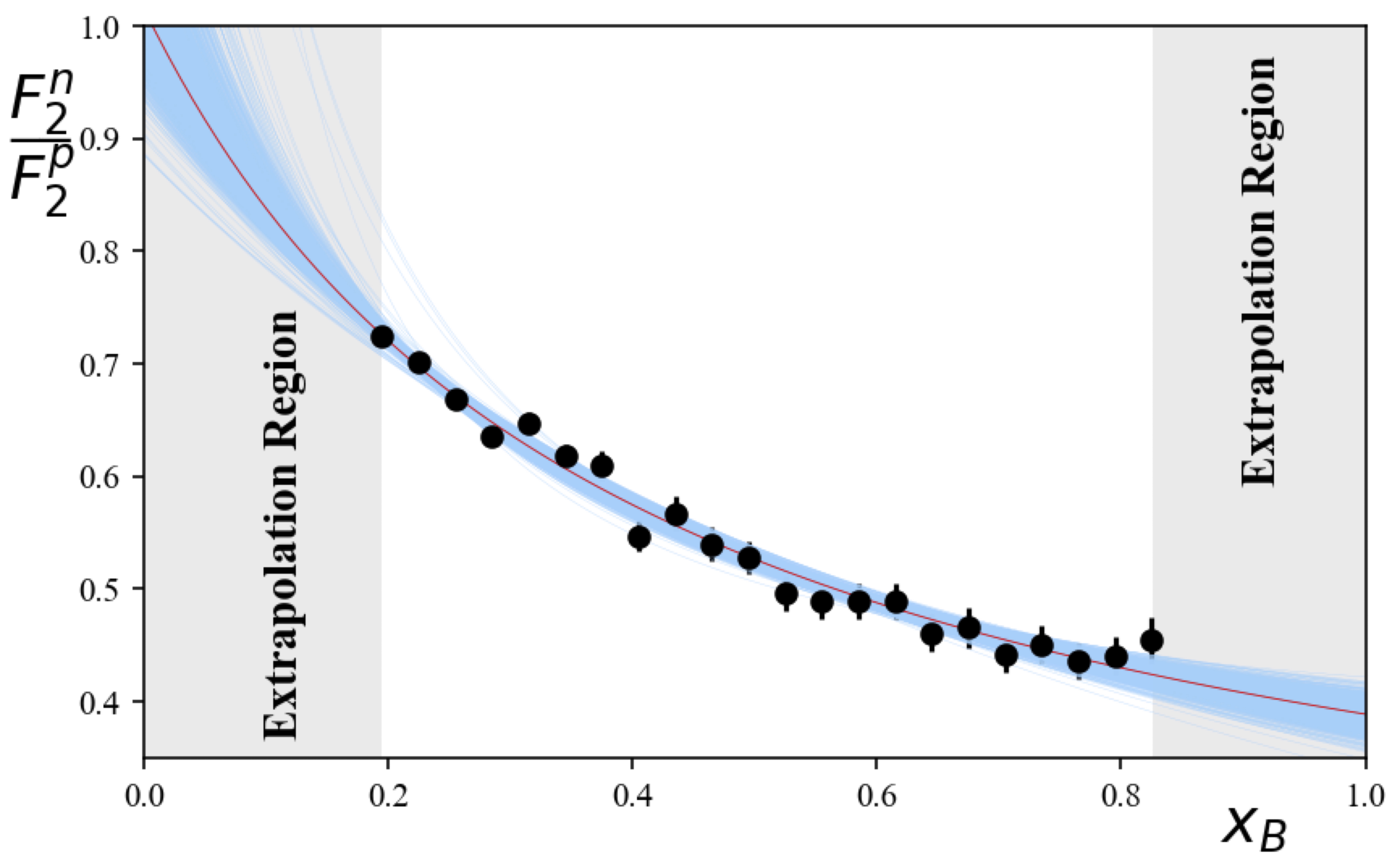}
    \caption{Fit of the Marathon data making use of a standard extrapolating function with the uncertainty band determined from statistical bootstrapping.   The result at $x_B=1$ indicates a ratio of 0.4 $\pm$ 0.05.}
    \label{stat-bootfinal}  
\end{figure}

\begin{figure*}[ht!]
    \includegraphics[width =0.45\textwidth, ]{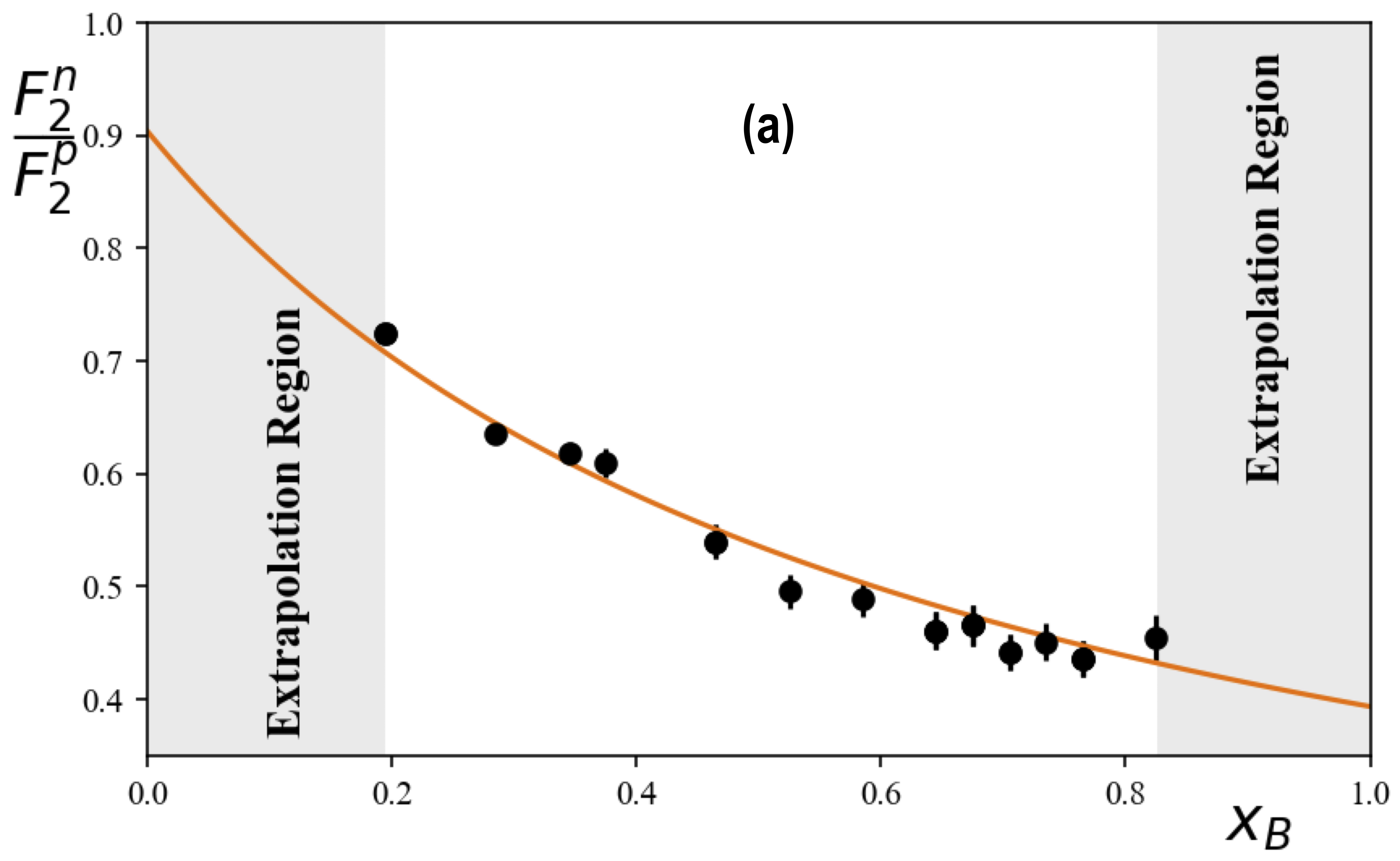}
\put (5,5) {\scriptsize a}
    \vspace{-0.555cm}
    \includegraphics[width =0.45\textwidth]{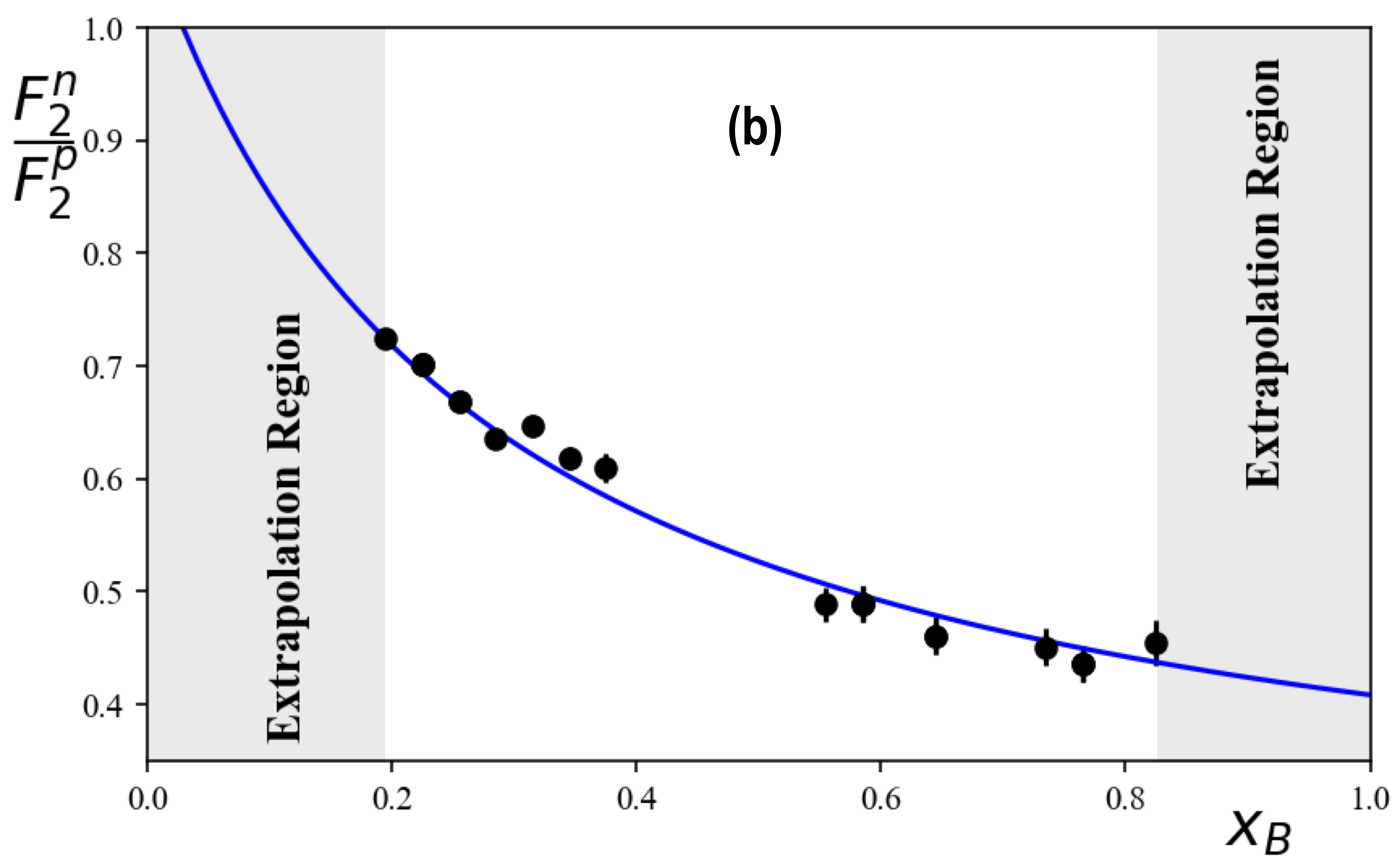}
    \includegraphics[width =0.45\textwidth]{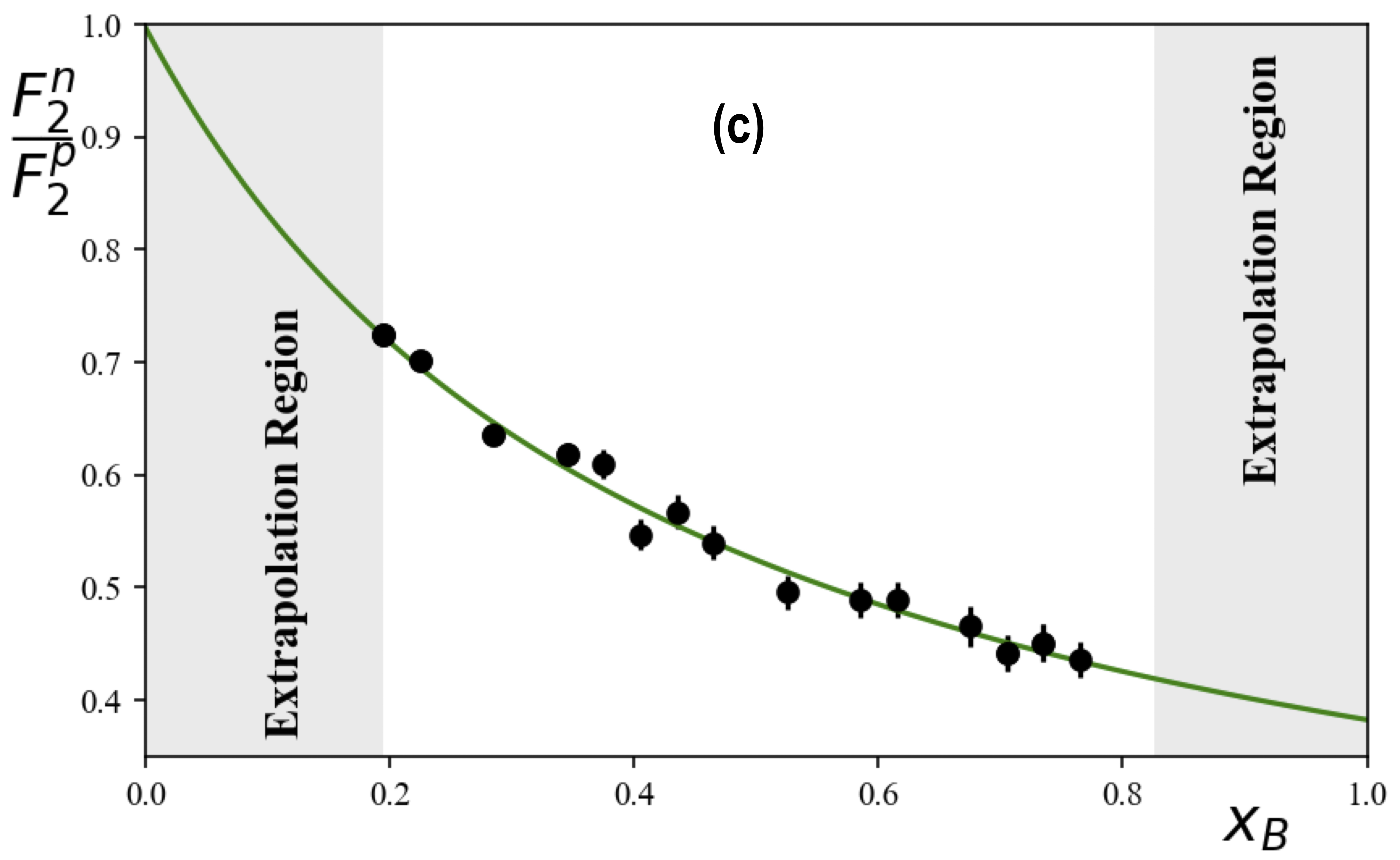}
     \vspace{-0.555cm}
    \includegraphics[width =0.45\textwidth]{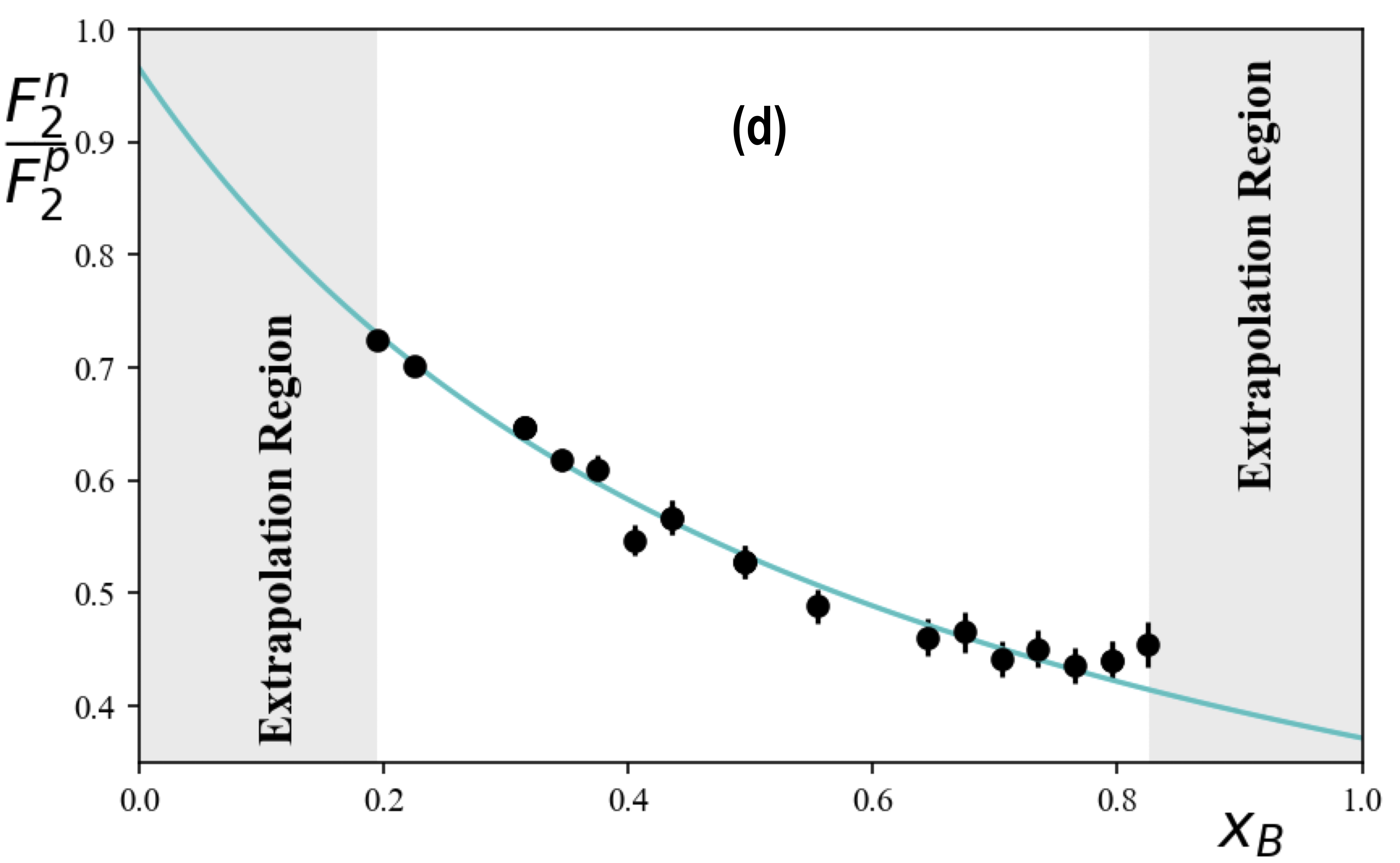}
    \includegraphics[width =0.45\textwidth]{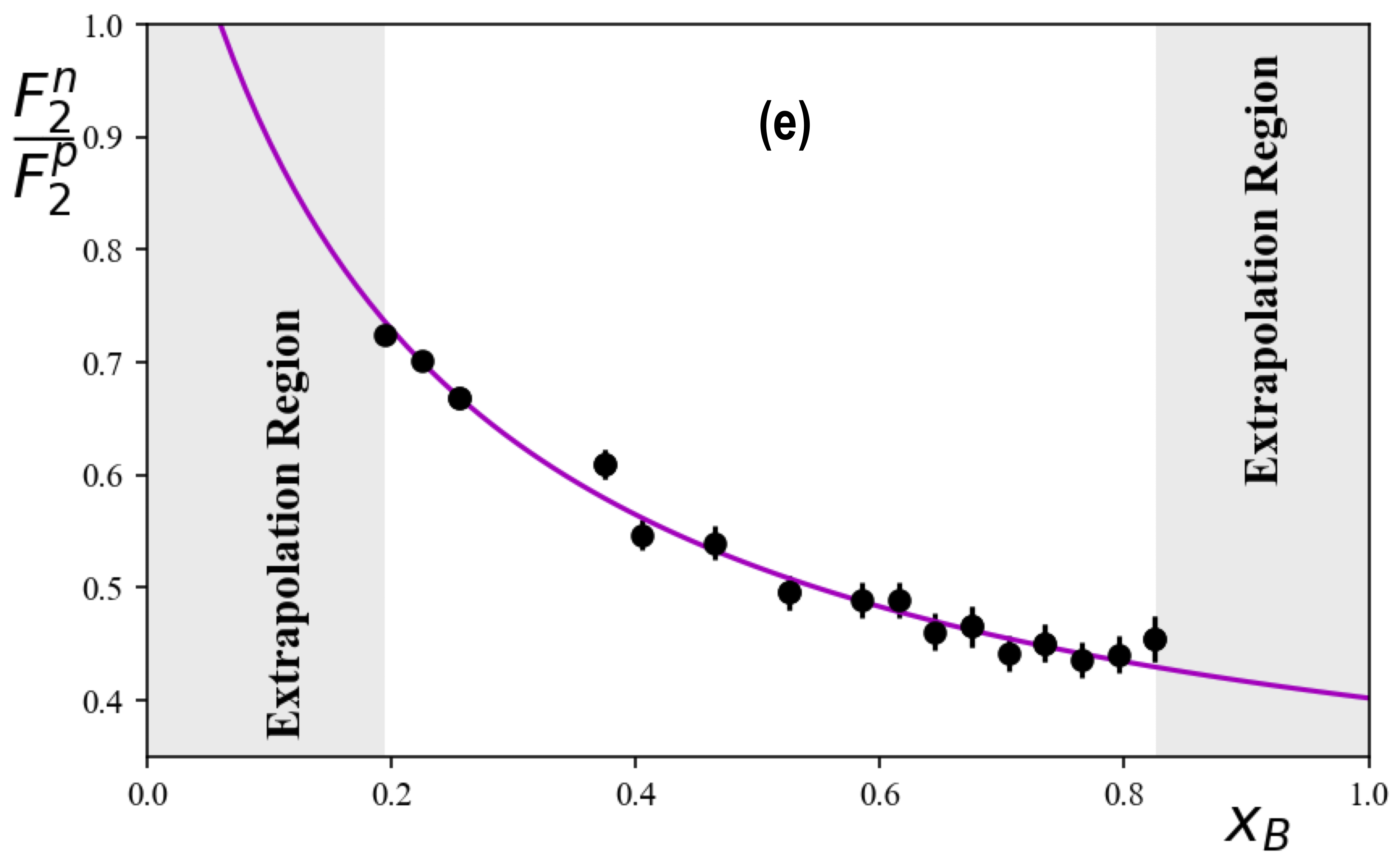}
          \vspace{-0.555cm}
    \includegraphics[width =0.45\textwidth]{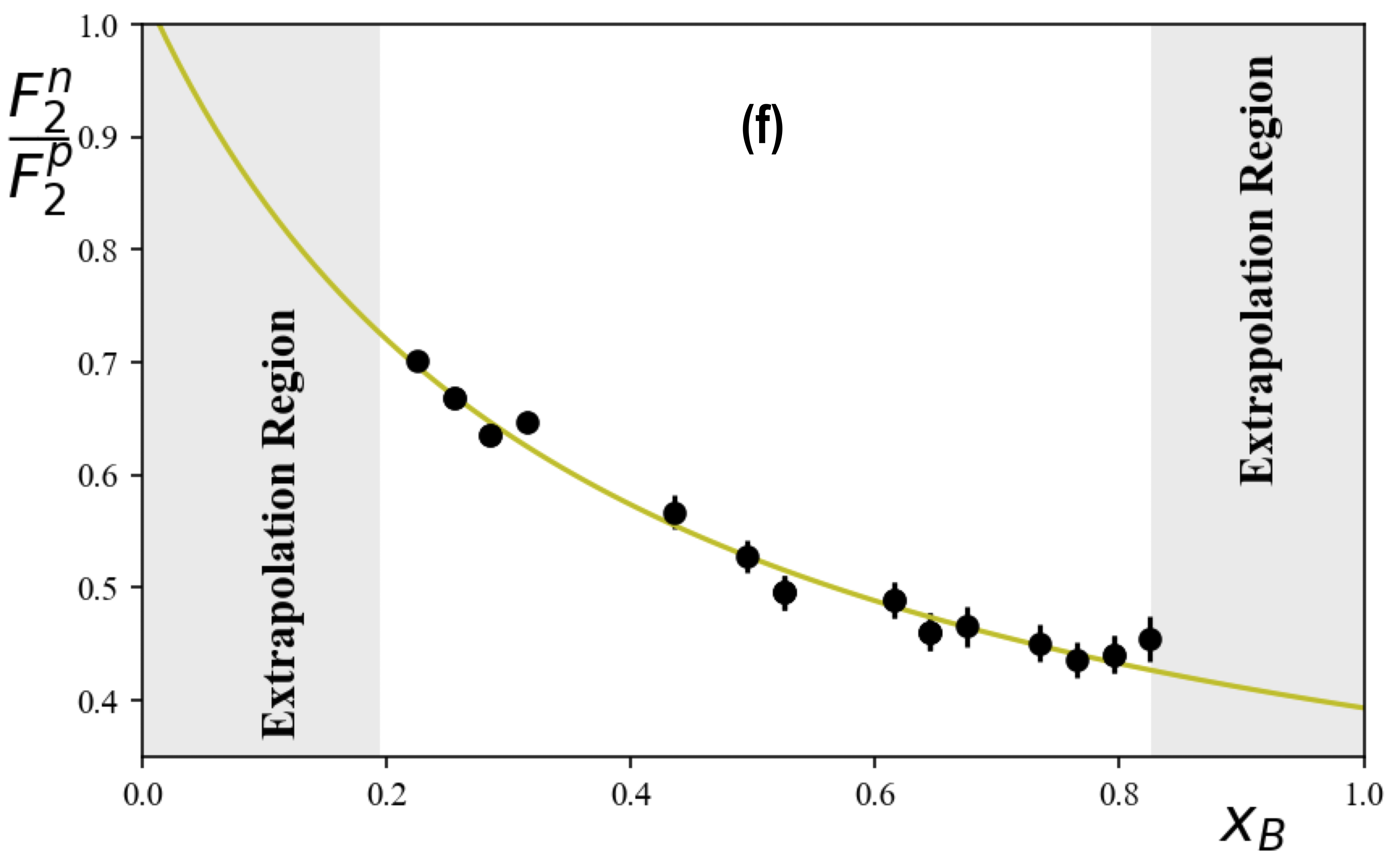} \includegraphics[width =0.45\textwidth]{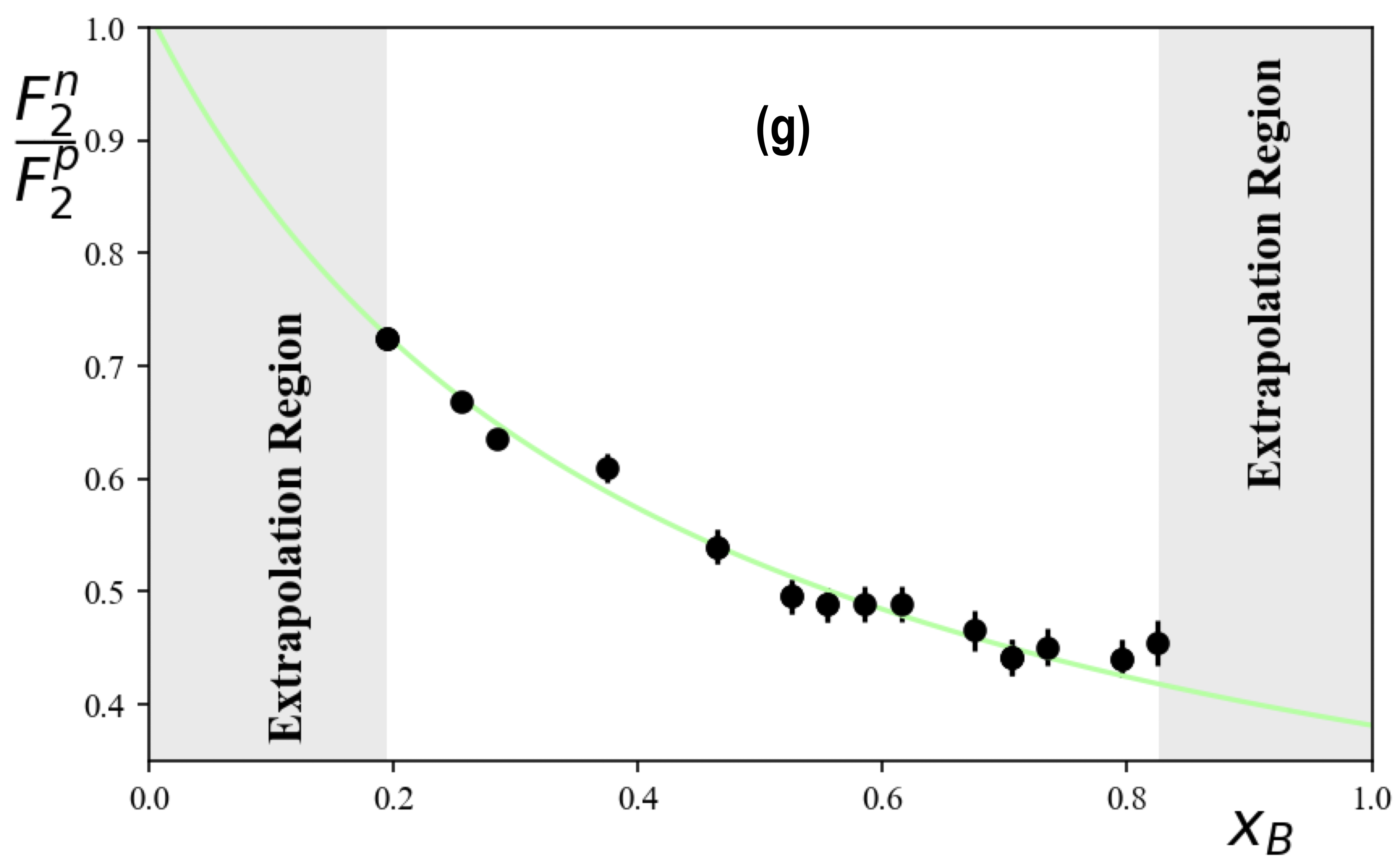}
    \vspace{-0.cm}
    \includegraphics[width =0.45\textwidth]{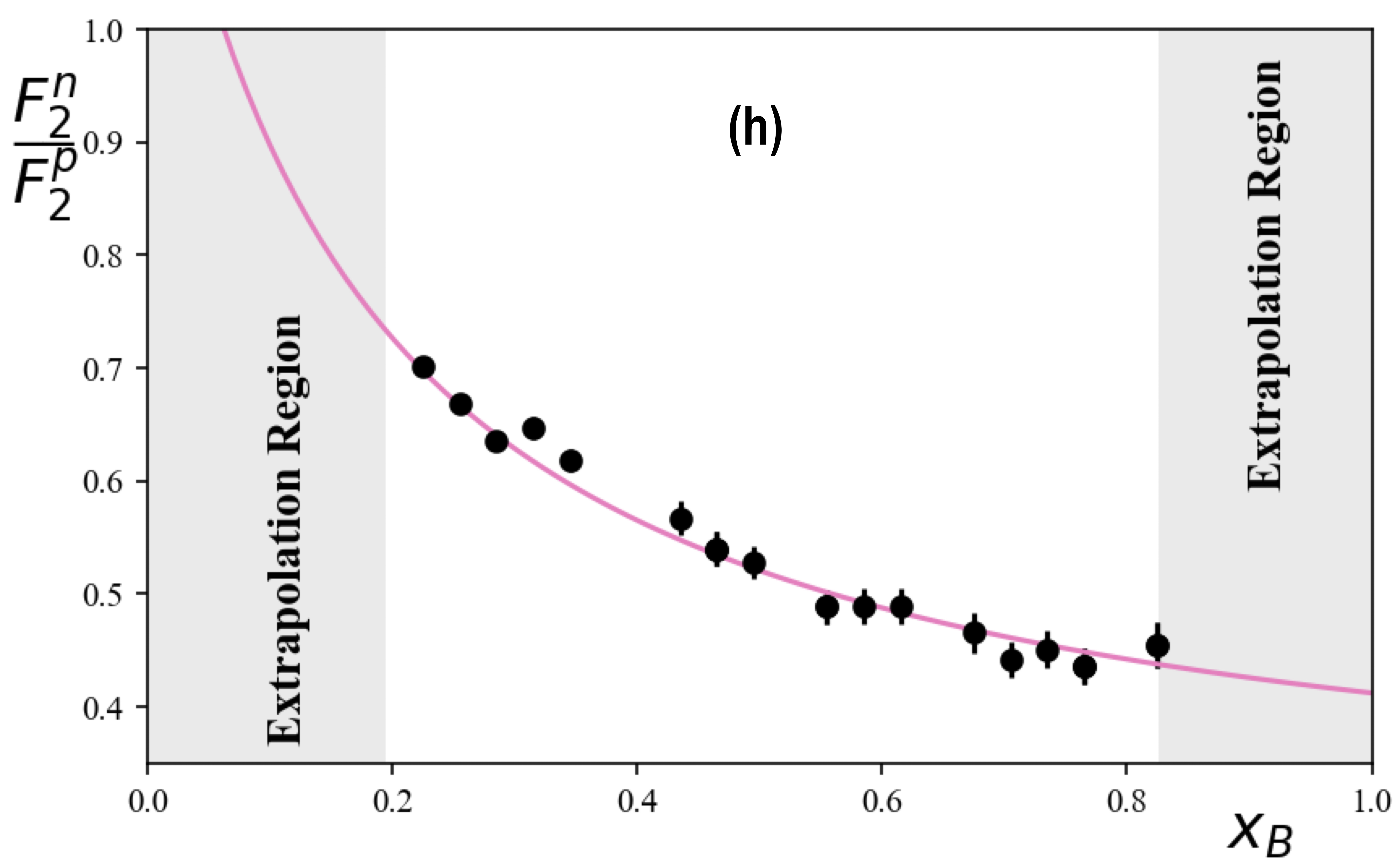}
    \caption{An 8-panel display of sample plots from the $\sim 10^3$ statistical bootstraps that were performed.   The method of uncertainty determination makes use of sampling with replacement and does not require that the data are normally distributed.}
    \label{stat-boot2}
\end{figure*}

 One fit line is created by randomly selecting a set of twenty-two points from the 22 existing data values allowing one, two or three substitutions of any of the data points with a randomly chosen value from the same data set. A total of one-thousand fits were performed using this method then plotted to determine the error band.  The fits were then extrapolated to lower and higher values of $x_B$ to cover the full range from  $x_B\rightarrow 0$ to $x_B\rightarrow 1$. Panels of selected fits using this statistical bootstrap method are presented in Fig.\ref{stat-boot2}. The blue error band shown in Fig.\ref{stat-bootfinal} is the result of the accumulation of one-thousand such fits.

\newpage

\section{Conclusion}
We have used a statistical bootstrap of the MARATHON deep inelastic scattering data to extrapolate the high and low quark momentum fraction regimes of the $F_2^n/F_2^P$ ratio.  As $x_B\rightarrow 1$, the ratio $F_2^n/F_2^P$ approaches $0.4 \pm 0.05$, which, under the usual isospin symmetry arguments for valence quark parton distribution functions of $u^p = d^n =u$ and $d^p=u^n=d$, corresponds to a ratio of $d/u \rightarrow 1/6$.  Our extrapolation favors the perturbative QCD helicity (pQCD) conservation, quark counting rules and DSE with contact interactions models.  The scalar diquark model value of the $x_B\rightarrow1$ limit $F_2^n/F_2^P=1/4$ is disfavored as are all other models from Table \ref{tab:predict}.  We note that while many models are incompatible with each other, it is possible that two compatible models could act over the same physical region to give the extrapolated value.  One example is the scalar diquark model with $F_2^n/F_2^P\rightarrow 0.25$ together with the 3-quark correlation plus MMHT2014NNLO with $F_2^n/F_2^P \geq 0.68$ from \cite{Leon:2020bvt}, with different weight given to the scalar diquark vs. 3-quark correlation contributions in order to yield a final ratio of $0.4$.  Such a scenario would agree with recent results on the nucleon wavefunction as a linear combination of a 3-valence quark state and a quark-diquark state with unequal coefficients in A=3 nuclei \cite{West:2020tyo,Li:2022fhh}.

\begin{acknowledgments}
We thank Wally Melnitchouk for helpful discussions and Tyler Kutz for his MARATHON plot.  H.V and F.B acknowledge support from the National Science Foundation, Award No.Benmokhtar-2012413. J.R.W is supported by the LDRD programs of LBNL, the EIC Center at Jefferson Lab and by the U.S. Department of Energy, Office of Science, Office of Nuclear Physics, under contract number DE-AC02-05CH11231.  This research was funded in part by Department of Energy grant number DE-AC05-06OR23177 
under which the Jefferson Science Associates operates the Thomas Jefferson National Accelerator Facility.
\end{acknowledgments}
\clearpage

\bibliography{boot-ratio}

\end{document}